\documentclass{emulateapj}
\usepackage{apjfonts}
\usepackage{lscape}

\newcommand \lsun{\hbox{$\hbox{L}_{\odot}$}} 
\newcommand \microjy{$\mu$Jy}
\newcommand \qlt{$q < 0$}
\newcommand \qprime{$q^{\prime}$}
\newcommand \qprimecorr{$q^{\prime}_{\rm corr}$}

\slugcomment{Accepted for publication in The Astrophysical Journal}
\shorttitle{X-ray Non-Detected AGN}
\shortauthors{DONLEY ET~AL}

\begin{document}

\title{Unveiling a Population of X-ray Non-Detected AGN}

\author{
J. L. Donley,\altaffilmark{1}
G. H. Rieke, \altaffilmark{1}
J. R. Rigby, \altaffilmark{1}
P. G. P\'{e}rez-Gonz\'{a}lez \altaffilmark{1}
}

\altaffiltext{1}{Steward Observatory, University of Arizona, 933 
North Cherry Avenue, Tucson, AZ 85721; jdonley@as.arizona.edu}


\begin{abstract}

We define a sample of 27 radio-excess AGN in the {\it Chandra} Deep
Field North by selecting galaxies that do not obey the radio/infrared
correlation for radio-quiet AGN and star-forming galaxies.
Approximately 60\% of these radio-excess AGN are X-ray undetected in
the 2~Ms {\it Chandra} catalog, even at exposures of $\ge$ 1~Ms; 25\%
lack even 2$\sigma$ X-ray detections. The absorbing columns to the
faint X-ray-detected objects are $10^{22}$ cm$^{-2} <$ N$_H$ $<
10^{24}$ cm$^{-2}$, i.e., they are obscured but unlikely to be Compton
thick. Using a local sample of radio-selected AGN, we show that a low
ratio of X-ray to radio emission, as seen in the X-ray weakly- and
non-detected samples, is correlated with the viewing angle of the
central engine, and therefore with obscuration. Our technique can
explore the proportion of obscured AGN in the distant Universe; the
results reported here for radio-excess objects are consistent with but
at the low end of the overall theoretical predictions for
Compton-thick objects.

\end{abstract}



\section{Introduction}

Hard X-ray surveys are an efficient means to define AGN samples, as
hard X-ray emission is able to penetrate the dust and gas capable of
obscuring an AGN's optical, UV, and soft X-ray emission. However, AGN
with high columns of absorbing gas can be missed even in the hard
(2-10~keV) X-ray.  Many local AGN have very high absorbing column;
Seyfert 2's are about four times more numerous than Seyfert 1s
(Maiolino \& Rieke 1995), and more than half of Seyfert 2 nuclei are
Compton thick with N$_{\rm H} > 10^{24}$~cm$^{-2}$ (Maiolino et
al. 1998; Risalti et al. 1999). X-ray background population synthesis
models predict that AGN are similarly obscured out to high ($z=2-3$)
redshift; $\sim $60\% of high-redshift AGN are predicted to have
obscuring columns greater than N$_{\rm H} = 10^{23}$~cm$^{-2}$ (e.g.,
Comastri et al. 2001, Gilli 2004). If distant AGN are indeed this
obscured, the deep hard X-ray {\it Chandra} surveys will be moderately
incomplete for obscured AGN (N$_{\rm H}>10^{22}$~cm$^{-2}$); Treister
et al. (2004) estimate 25\% incompleteness at N$_{\rm H} = 3 \times
10^{23}$~cm$^{-2}$. Such surveys will also miss nearly all of the
Compton-thick AGN, a large population of which are expected both
locally and in the high-redshift Universe (c.f. Ueda et al. 2003).

Quantifying the fraction of AGN missed by current surveys is necessary
if we are to understand the accretion history of the Universe.  In
addition, the identification of heavily obscured AGN is in itself
interesting, as such sources are poorly understood in the distant
Universe.

The availability of deep mid-infrared (MIR) and radio images of the
{\it Chandra} Deep Field North (CDFN) provides a new way to search for
heavily obscured AGN, as extinction in the MIR and radio is
small. Radio-excess AGN are significantly brighter in the radio
relative to the infrared than are star-forming galaxies, which fit the
well-defined radio/infrared correlation (Helou et al. 1985; Appleton
et al.  2004).  With the deep MIR data, we have separated the
radio-excess AGN population from star-forming galaxies by comparing
their radio and MIR flux densities. Because this radio/infrared method
of classification is independent of the optical, UV, and X-ray
characteristics of the deep radio sample, we detect via this selection
approach AGN that are missed at other wavelengths, thereby testing the
completeness of current AGN samples.  We find that a significant
number of radio-excess AGN do not appear even in the deepest available
X-ray surveys.

This paper is organized as follows. In \S2 we discuss the radio and
infrared observations and data reduction. The X-ray faint AGN
population is introduced in \S3, followed by a discussion of redshifts
in \S4. We focus in \S5 on lower-significance X-ray detections, along
with the X-ray coaddition of undetected AGN.  The discussion follows
in \S6, and we summarize the conclusions of the paper in \S7.


\section{Observations and Data Reduction}

Our initial sample consists of 371 radio sources detected at 1.4~GHz
by the Richards et al.~(2000) VLA survey of the CDFN; this survey has
a detection limit of 40~$\mu$Jy, rms noise of $\sim$ 7.5~$\mu$Jy, and
is 95\% complete to 80~\microjy\footnote{A rereduction of the radio
data by G. Morrison shows flux density differences from the original
reduction that are generally no more than $\sim$ 10\%, with the
notable exception of source VLA J123659+621832, where there is a
discrepancy by a factor of ten (private communication, F. Bauer and
G. Morrision). Adopting the new value for this source would push it
further into the AGN regime in our selection method but would have no
effect on the conclusions of this paper.}. We later restrict this
initial sample to those sources with X-ray exposures $>$ 1~Ms; this
reduces the radio sample size to 162, and the effective area of the
survey to $\sim~$230~arcmin$^2$ (Alexander et al. 2003).

The 24 \micron\ properties of the radio sources are determined from a
$\sim$ 1400s Spitzer MIPS observation of the CDFN.  The MIPS
24~\micron\ image was processed with the MIPS GTO data analysis tool
(Gordon et al. 2004).  The measured count rates are corrected for dark
current, cosmic-rays, and flux nonlinearities, and are then divided by
flat-fields for each MIPS scan-mirror position. Images are then
interpolated to half pixels, corrected for geometric distortion, and
mosaicked.  The final mosaic has a pixel scale of
$1^{\prime\prime}\!\!.25$ pix$^{-1}$, with a point-spread function
full-width at half maximum (FWHM) of $\approx 6^{\prime\prime}$.
Source extraction at 24~\micron\ was performed using the task
\textit{allstar} in the DAOPHOT package of IRAF\footnote{IRAF is
distributed by the National Optical Astronomy Observatories, which is
operated by the Association of Universities for Research in Astronomy,
Inc. (AURA) under cooperative agreement with the National Science
Foundation} (see P\'{e}rez-Gonz\'{a}lez et al.~2005). We detect 1868
24 \micron\ sources in the $>$~1~Ms field, 1167 of which have flux
densities higher than the 80\% completeness limit of the MIPS
24~\micron\ image processed in this manner, 80~\microjy\ (Papovich et
al. 2004). Systematic photometric errors in the MIPS photometry are
small ($\sim 1\%$), so the measurement accuracy is determined by
statistical errors (about 15$\mu$Jy rms) and confusion noise (about
11$\mu$Jy rms).

The 1.4~GHz and 24~\micron\ sources were matched using a 2\arcsec\
search radius, after adjusting the 24~\micron\ positions to remove any
systematic positional offset between the radio and infrared images. We
exclude from our final radio-excess AGN sample one radio source whose
nearest 24~\micron\ counterpart lies between 2\arcsec\ and 3\arcsec\
of the radio position, as the reliability of counterparts in this
region is ambiguous.  If a source was not detected because it is
blended with a brighter 24~\micron\ source, we use the flux density of
the blended source as a conservative upper limit.  Sources that were
not detected at 24~\micron\ were assigned the $\sim 5\sigma$ limit of
80~\microjy. At the detection limits, there are $\sim$ 0.7 radio
sources and $\sim$ 5 infrared sources per square arcmin. Consequently,
one expects $\sim$ 3 false coincidences of radio and 24$\mu$m sources
in the entire sample. Since we identify AGN by their relatively faint
24$\mu$m flux densities, the false detections will tend to remove some
AGN from our sample but will not bias our results otherwise.

We also use a 500~s IRAC observation of the CDFN. The IRAC
observations were divided into 4 dithers of 125~s each; the 5$\sigma$
point source sensitivity limits for the IRAC 3.0, 4.5, 5.8, and
8.0~\micron\ bands are 1.4, 2.8, 18.4, and 22.7~\microjy\ respectively
(Fazio et al.~2004). The IRAC data analysis is described in Huang et
al. (2004). Source extraction in the IRAC bands was conducted using
SExtractor (Bertin \& Arnouts 1996).

Optical and near-infrared (NIR) photometry are measured from datasets
published for the CDFN by the GOODS team (\textit{bviz}, Giavalisco et
al. 2004) and by Capak et al. (2004) (\textit{UBVRIzHK}). When there
is a single optical counterpart within 2\arcsec\ of the radio source,
the optical and NIR magnitudes are aperture-matched for consistency as
described in P\'{e}rez-Gonz\'{a}lez et al.~(2005). For radio sources
with more than 1 optical or NIR counterpart within a 2\arcsec\ radius
of the radio position, we choose the counterpart that is nearest the
radio position and use its individual SExtractor magnitude, measured
within an elliptical aperture corresponding to 2.5 times the Kron
radius (Kron 1980). The magnitudes measured in the GOODS dataset in
this way are identical to those given by Giavalisco et
al. (2004). When multiple GOODS counterparts are available, we choose
the nearest source and do not use the ground-based optical data, which
is likely to be blended.


\begin{figure*}[t]
\epsscale{0.8}
\plotone{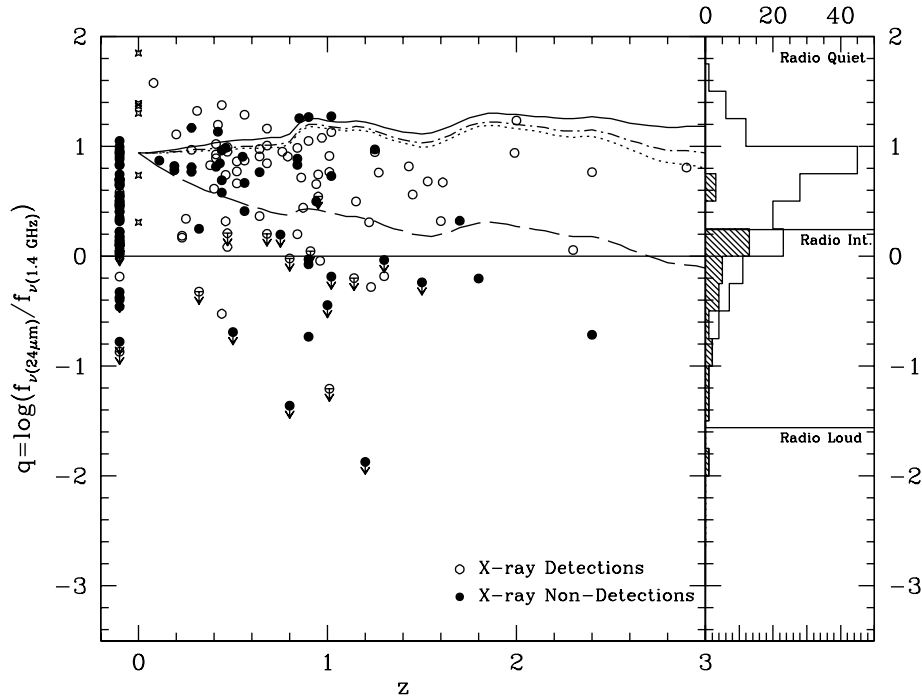}
\caption{Relationship between $q$ and redshift.  All sources without 
known redshifts have been assigned $z = -0.1$. Circles represent all
radio sources with X-ray exposures in excess of 1~Ms.  Crosses
represent the 6 lowest-metallicity galaxies from the Engelbracht et
al. (2005) sample. The lines represent the change in the observed $q$
with redshift for SEDs with LINER (solid), Seyfert 2 (dot-dashed),
spiral galaxy (dotted), or ULIRG (dashed) templates, normalized to the
best-fit local value of $q$ as determined by Appleton et
al. (2004). The histogram shows the distributions of $q$ in the
radio-quiet, radio-intermediate, and radio-loud regimes, where the
shaded histogram represents sources for which we have only upper
limits on $q$, and the clear histogram gives the distribution of $q$
for all sources.  We define as our radio-excess AGN sample all
galaxies with \qlt.}
\end{figure*}

\section{The X-ray Faint AGN Population}

We use the 1.4~GHz and 24~\micron\ flux densities of our sample, along
with the well-known radio/infrared relation of star-forming galaxies
and radio-quiet AGN, to select a sample of radio-excess AGN
independently of their X-ray properties.  Appleton et al.~(2004)
define $q$ = log (f$_{24~\micron}$/f$_{1.4~\rm GHz}$). They show for
star-forming galaxies that $q$ is tightly constrained out to z $\sim$
1$:$ $q$ = 0.94 $\pm$ 0.23 after K-correction. Galaxies with values of
$q$ well below this range (strong radio with respect to 24~\micron\
emission) are unlikely to be dominated by star formation and instead
are radio-emitting AGN (Drake et al. 2003).

\vspace{1cm}

\subsection {Sample Selection}

Figure 1 illustrates the selection criteria. We set a selection
threshold of \qlt\ to classify a galaxy as probably having an AGN and
require a {\it Chandra} exposure of $>$ 1~Ms to provide high X-ray
sensitivity. This selection identifies 27 radio-excess AGN in the
CDFN, as listed in Table 1. Of these, 16 (60\%) have no X-ray
counterpart in the Alexander et al. (2003) catalog.

Because of the lack of spectral information, we have chosen not to
K-correct our radio and infrared flux densities and instead illustrate
in Figure 1 the redshift evolution of $q$ for galaxies with LINER,
Seyfert 2, spiral galaxy, and ultra-luminous infrared galaxy (ULIRG)
templates, normalized to the best-fit value of Appleton et al.~(2004).
Templates were taken from Devriendt et al. (1999). A spectral index of
$\alpha = 0.7$ ($S_{\rm\nu} \propto \nu^{-\alpha}$) was assumed for
the radio continuum.  For the LINER, Seyfert 2, and spiral templates,
the observed $q$ is slightly larger than the intrinsic value out to
redshifts of $z \ge 2.7$.  Redshifting a ULIRG SED causes the observed
$q$ to drop below the intrinsic value as $z$ increases.

Because we have not K-corrected our flux densities, we use the
slightly higher non K-corrected dispersion, $\sigma = 0.28$, from
Appleton et al. to calculate the significance of our selection
threshold at $z=0$.  Our selection threshold of \qlt\ is 3.4 standard
deviations below the average value of $q$ at $z=0$, and therefore our
sample should not be significantly contaminated with normal
star-forming galaxies. The significance of our selection at $z > 0$
depends on each object's individual SED, and will increase slightly
for LINERS, Seyfert 2's and spirals, but decrease for ULIRGS. We
estimate the maximum error on the measured values of $q$ by assuming
both a radio and infrared flux density of 80 \microjy, the approximate
limits of our survey.  This corresponds to a 5$\sigma$ 24 \micron\
detection and a 10$\sigma$ radio detection and gives an error of 0.1
on $q$; we therefore do not expect our sample to be strongly affected
by errors in the flux determination. The crowded nature of several of
the fields implies that some of our sources may have blended
24~\micron\ emission. This will cause $q$ to increase, however, and
will therefore remove sources from our radio excess sample instead of
introducing contaminants.

Yun et al. (2001) define radio-excess objects as those with radio
emission exceeding the radio/FIR correlation prediction by a factor of
$\ge$~5 (a behavior hereafter described as a radio excess by a factor
of 5).  From the measurement of this correlation by Appleton et
al. (2004), the corresponding threshold would be $q$ $<$ 0.24 at $z =
0$. At $z = 1$, the radio-excess threshold ranges from $q = 0.46$ to
$q = 0.53$ for LINER, Seyfert 2, or spiral galaxy templates. For all
these cases, our selection criterion of \qlt\ (which corresponds to a
radio excess of 8.7) isolates radio-excess AGN.  For a ULIRG template,
the radio-excess threshold is -0.30 at $z=1$; a $q$ of 0 for such a
template at $z=1$ corresponds to a radio excess of 2.5, half the AGN
threshold suggested by Yun et al. (2001).  As will be discussed in
\S4, however, a small fraction of the sources in our sample have ULIRG
SEDS, and all are likely to be AGN.

Following Drake et al. (2003), radio-loud objects would have $q <
-1.6$ at $z = 0$; since few sources fall in this range, the AGN sample
consists almost entirely of radio-intermediate sources, as illustrated
in the histogram inset in Figure 1.

\subsection{Potential Sources of Contamination}

At high redshift, the 24~\micron\ emission of normal star-forming
galaxies is dominated by PAH features, which are absent at low
metallicity (e.g., Houck et al. 2004, Engelbracht et al. 2005). As
Figure 1 demonstrates, however, low metallicity star-forming galaxies
also tend to have $q > 0$ and therefore they should not contaminate
our radio-excess AGN sample.

Nearly all star-forming galaxies in the field fit the radio/infrared
correlation. Even within dense galaxy clusters, where environmental
effects might boost the radio emission, the relation holds well with
only a tiny minority of galaxies having radio excesses as large as a
factor of five (Miller \& Owen 2001). Bressan et al. (2002) show that
a radio excess is expected in post-starburst galaxies due to the
slower fading time of the radio emission with respect to the FIR;
their models predict a maximum radio enhancement of $\sim 5$ above the
radio/FIR correlation. As mentioned above, our selection criterion
selects AGN whose radio enhancements exceed a factor of $\sim 9$ in
LINER, Seyfert 2, and spiral galaxy templates. We therefore expect
minimal contamination from post-starburst galaxies.

The strong silicate absorption feature seen in a handful of ULIRGS is
another potential source of sample contamination. This feature is
centered at 9.7~\micron\, and would therefore be redshifted into our
24~\micron\ band at redshifts of $z \sim 1.2-1.5$, where it could
suppress the observed infrared flux density. Four of the X-ray
non-detected radio-excess sample have redshifts in this range.

It is unlikely, however, that a large fraction of AGN are affected in
this way. The 12~\micron\ flux of AGN provides the best estimate of
their bolometric luminosity, allowing for effective infrared selection
(Spinoglio \& Malkan 1989).  The silicate absorption feature is within
the IRAS 12~\micron\ passband, and evidently has only a secondary
effect on the luminosity estimate.  In fact, $\sim 50$\% of
high-luminosity AGN have silicate {\it emission} features (Hao et
al. 2005). In such cases, the $q$ of true AGN will be shifted towards
the star-forming range, resulting in incompleteness in our sample, but
not contamination.

\subsection{Sample Completeness}

Because the 95\% completeness limit of the radio sample is equal to
the upper limit we can place on the flux densities of non-detected 24
\micron\ sources, 80~\microjy, restricting the radio flux densities to
$S_{\nu} > 80$~\microjy\ would provide a 95\% complete radio-excess
(\qlt) sample (subject to the cautions mentioned above). While we do
not apply a radio flux-density cut for the sample discussed in this
paper, only one AGN in our sample, an X-ray non-detected source, has a
radio flux density that falls below this limit. In addition, the
counterpart uncertainties and 24 \micron\ blending discussed in \S2
only led to the removal of one source that could fall into such a
complete sample. Therefore, if we were to define a complete sample by
requiring that $S_{\nu} > 80$~\microjy, our sample would essentially
remain the same. Also, in \S2 we estimated that up to 3 radio-excess
AGN would be removed from our sample due to chance coincidences with
24$\mu$m sources; this process will reduce the sample size but have no
other effect. Thus, we can say that for a sample of 27 radio-excess
objects in the CDFN (with X-ray exposures in excess of 1~Ms) that is
sufficiently complete to be largely unbiased for this class of AGN, 16
(60\%) sources are previously uncataloged as AGN. The remainder of
this paper probes the characteristics of these objects in more detail.


\section{Redshifts}

Spectroscopic redshifts for 8 AGN, and photometric redshifts for an
additional 2, were taken from the literature (Barger et al. 2002,
Cowie et al. 2004, Wirth et al.  2004 (TKRS)).  The redshifts are
tabulated in Table 1 and Figure 1. All but one of the AGN for which
Team Keck Redshift Survey (TKRS) spectra are available appear to be
early-type galaxies with no obvious signs of nuclear activity in the
optical/UV. The remaining source, VLA~123646+621404, is X-ray detected
and has weak narrow emission lines.

For most of the remaining AGN in our sample, we estimated redshifts
photometrically by fitting the Devriendt et al. (1999) templates to
the galaxy optical through MIR spectral energy distributions (SEDs).
SED fitting was conducted both by an automated fitting routine and by
eye; the two methods give results consistent to within the errors
quoted below. The SEDs of all radio-excess AGN in our sample are shown
in Figure 2. P\'{e}rez-Gonz\'{a}lez et al.~(2005) provide further
details on use of Spitzer and other data to obtain photometric
redshifts. The redshifted SED fits are in good agreement with
spectroscopic redshifts in all cases where spectroscopic data are
available. We are able to find or estimate redshifts for 12 of the 16
X-ray non-detected AGN and for 8 of the 11 X-ray detected AGN. The
typical errors in our photometric redshift determinations are
$\Delta(z)/(1+z) < 0.1$ (see P\'{e}rez-Gonz\'{a}lez et al.~2005).  The
mean redshifts of the X-ray non-detected and X-ray detected AGN
samples are $z = 1.2$ and $z = 0.9$, with dispersions of 0.5 and 0.4,
respectively.

\begin{figure*}[p]
\centering
\epsscale{1.1}
\plotone{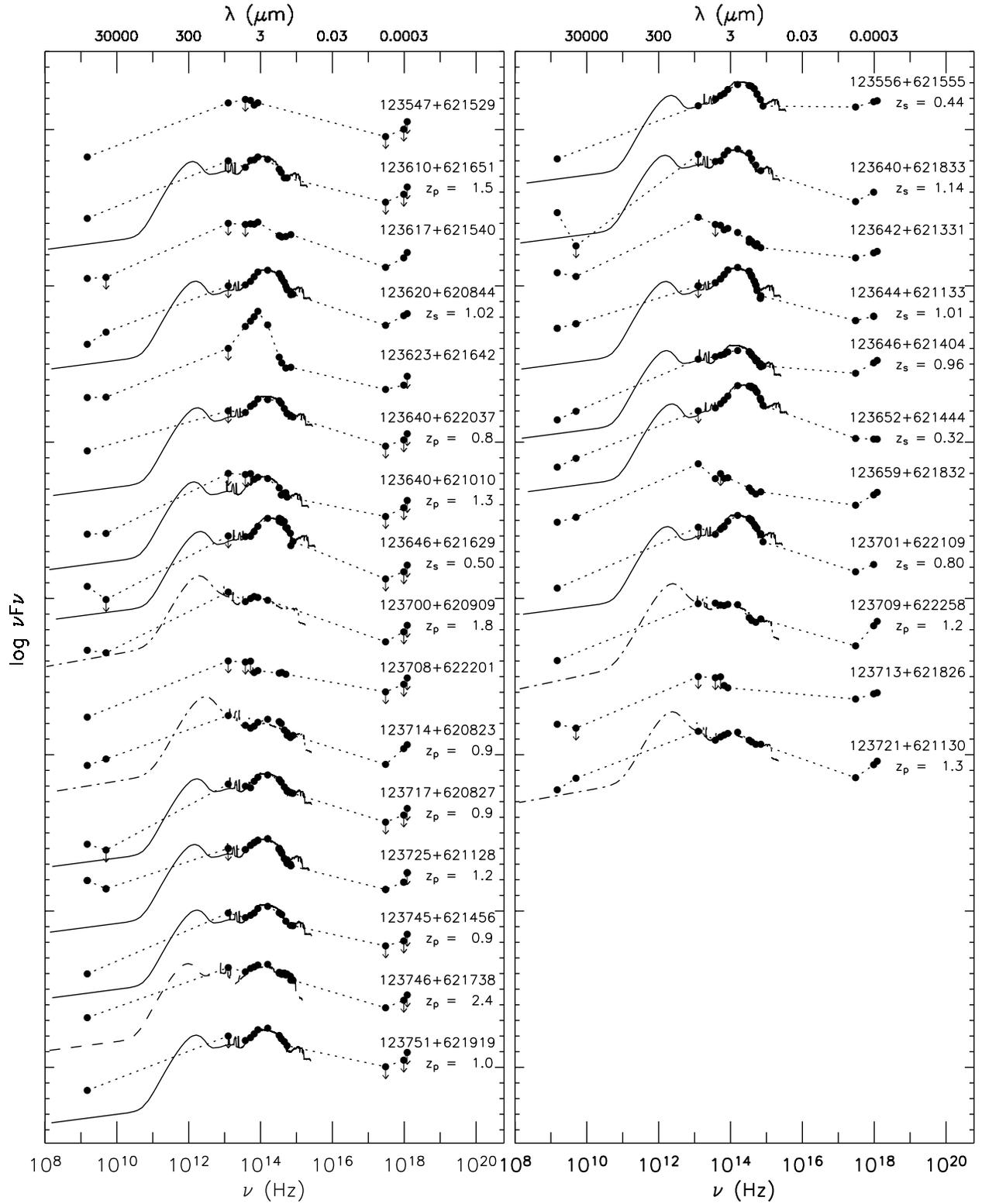}
\caption{Observed-frame SEDs of the radio-excess AGN. Sources with weak
or non-detected X-rays are plotted in the left panel, and sources with
high-significance X-ray detections are plotted in the right panel.  We
list VLA source IDs and photometric (z$_{\rm p}$) or spectroscopic
(z$_{\rm s}$) redshifts. Templates were taken from Devriendt et
al. (1999). Seyfert 2 and LINER templates are given by solid lines.
Dot-dashed lines represent ULIRG templates, and dashed lines represent
spiral templates. Large ticks along the y-axis represent a interval of
10 in log $\nu F_{\nu}$.}
\end{figure*}

Source VLA~J123642+621331 has a published redshift of $z=4.42$, based
on a single emission line believed to be Ly$\alpha$ (Waddington et
al. 1999).  The validity of this measurement has been questioned by
Barger et al. (2000). In addition, none of the Devriendt et
al. templates are able to fit the global SED, so we do not assign a
redshift to this source.

The Devriendt et al. templates used to obtain photometric redshifts
are comprised of spirals, LINERS, Seyfert 2's, starbursts, and ULIRGS.
Of the X-ray non-detected AGN, 9 are best-fit by LINER or Seyfert 2
templates, 2 have ULIRG templates, and 1 appears to be a spiral
galaxy.  Six of the X-ray detected AGN are best-fit by LINER or
Seyfert 2 templates; the remaining 2 are ULIRGS.

With redshifts and fitted SED templates for most of the sample, we can
test the robustness of our original $q < 0$ selection. Using the
best-fit Devriendt et al. SED templates and the radio spectral slopes
from Richards et al. (2000), we estimate the K-corrected radio
excesses for the 20 sources with estimated redshifts. All but 1 of the
X-ray non-detected AGN (with known redshifts) have radio excesses well
above both the Yun et al. ``radio-excess'' definition of 5 and our
$q<0$ selection, which is equivalent to a radio excess of $\sim
9$. The remaining source has a radio excess of 4 and thus should also
be dominated by AGN radio emission.  Similarly, 6 of the 8 X-ray
detected AGN with known redshifts have radio excesses greater than 10;
an additional source has a radio excess of 4, and therefore is likely
to be AGN-dominated.  The remaining source has a radio excess of only
1.3.  This source, VLA~123721+621130, is best-fit by a ULIRG template,
has an inverted radio spectrum, $\alpha = -0.28$, a very hard X-ray
photon index, $\Gamma = 0.28$, and an absorbed X-ray luminosity of
log~$ L_{0.5-8\ \rm{keV}} = 42.7$, all of which indicate the presence
of an AGN.  We therefore do not remove it from our sample.


\vspace{1cm}
\section{X-ray Detection and Coaddition}

Although the X-ray non-detected AGN all fall below the Alexander et
al. (2003) {\it Chandra} detection limit, detection to lower
significance is possible now that source positions are known \textit{a
priori}. We test for X-ray emission by comparing the X-ray counts
measured at the positions of our radio sources to the local X-ray
background distributions. Four of the X-ray non-detected AGN were
excluded from this analysis, as they are separated from known X-ray
sources by less than 2 times their 95\% encircled energy radii (EER),
and are therefore likely to be contaminated. We find signals for 7 of
the 12 remaining X-ray ``non-detected'' AGN at significance levels of
2.5 to 5.2 $\sigma$. For the remaining 5 sources, we use a stacking
analysis to constrain the average flux. We give details of these
measurements below.

\subsection {X-ray Detection}

To search for X-ray signal from each of the X-ray non-detected AGN, we
first extract the total number of X-ray counts within the 60\%, 70\%,
and 80\% EER centered on the radio position, using the CDFN 2~Ms image
provided by Alexander et al. (2003). The average source exposure time
is similarly extracted from the X-ray exposure maps. The EER for the
full (0.5-8~keV), hard (2-8~keV), and soft (0.5-2~keV) X-ray bands
were calculated using the {\sc mkpsf} task in CIAO\footnote{{\it
Chandra} Interactive Analysis of Observations (CIAO),
http://cxc.harvard.edu/ciao/}, from which we created an
exposure-weighted average PSF for each source, taking into account the
sources' positions on the ACIS-I detector during each of the 20 CDFN
observations\footnote{At intermediate to large off-axis angles, the
CXC PSF libraries crop a small fraction (5-10\%) of the X-ray flux,
affecting the determination of the EER.  We do not attempt to correct
for this effect, as the PSFs of the weakly-detected sources are
undersampled.}. We assume a photon index of $\Gamma = 1.4$ when
calculating the EER, though we have verified that changing this to
$\Gamma = 1.0$ negligibly affects the results.

We use a Monte-Carlo simulation similar to that of Brandt et
al. (2001) to estimate the local sky background in the 3 X-ray bands
and 3 apertures.  For each source, we first choose a random position
that (1) lies within 1\arcmin\ of the radio source, (2) does not
overlap with the source aperture, (3) is a minimum distance of 2 times
the source's 95\% EER from any cataloged X-ray source, and (4) has an
exposure time that differs from the source exposure time by less than
50\%. The final condition is set to minimize the effect of large
gradients in exposure on our local sky background distributions. To
determine the sky background distributions, we measure the sky counts
at this random position using the 3 source apertures. We repeat 10,000
times.

Before fitting the sky distributions for each band and EER, we
normalize the counts to the source exposure time. We then fit a
Poisson distribution to the resulting distributions of normalized
background counts, from which we measure the mean sky background.  The
dispersion of the sky background determines the likelihood of
measuring by chance the observed number of counts in an aperture
centered on the radio source. We define as ``weakly-detected'' all
sources at $>2 \sigma$ above the background in at least 1 of the 3
{\it Chandra} energy bands, and in at least 1 of the 3
apertures. Although we require only a $>2 \sigma$ detection, all 7 of
the weakly-detected sources are at $> 2.5 \sigma$ in at least one of
the 3 bands, and 2 are detected to $\ge 5 \sigma$. In Table 2, we give
the X-ray properties of all X-ray weakly-detected AGN. For each source
and energy band, we list the results for the aperture (60\% EER, 70\%
EER, or 80\% EER) with the highest detection significance.  To
illustrate our detection method, we show in Figure 3 the X-ray
background distribution for VLA~J123617+621450, along with the source
signal, which corresponds to a 3$\sigma$ detection.

\begin{figure}[t]
\epsscale{1.1}
\plotone{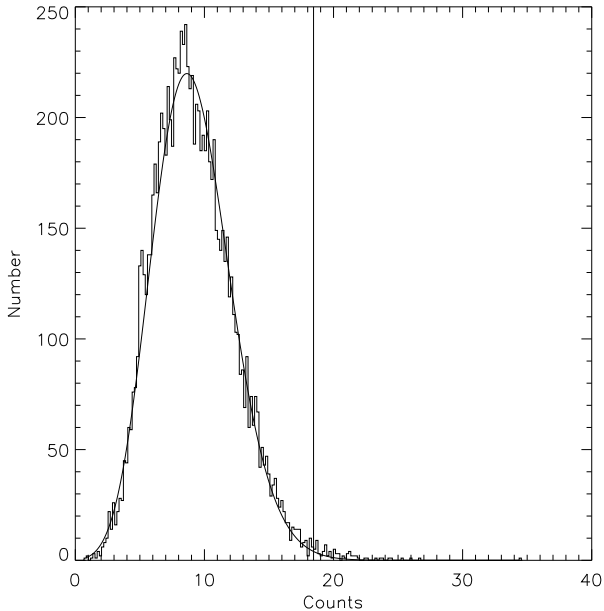}
\caption{X-ray background distribution for source VLA~J123617+621540 in 
the full X-ray band (0.5-8~keV).  The counts have been normalized to
the source exposure time.  The Poisson fit to the background
distribution is shown. The vertical line indicates the total number of
counts obtained at the source position, and corresponds to a 3$\sigma$
detection. }
\end{figure}

To calculate the X-ray flux from the aperture-corrected count rate, we
first need to know the X-ray photon index, $\Gamma$. If a given source
is detected to $\ge 2 \sigma$ in both the hard and soft X-ray bands,
the photon index is estimated from the ratio of the hard to soft count
rates, $H/S$. If a source is not detected to $\ge 2 \sigma$ in both
the hard and soft bands, we assume that $\Gamma = 1.4$. Although
$\Gamma$ spans a fairly wide range of values for the sources where it
can be estimated, the derived fluxes should only be weakly dependent
on the value adopted. For consistency, we use the {\it Chandra} 2 MS
Catalog $H/S$ to $\Gamma$ and X-ray count rate to X-ray flux
conversions (Alexander et al. 2003).  For comparison with the
cataloged fluxes, we do not correct the X-ray fluxes for Galactic
absorption.  The quoted X-ray flux ratios, however, have been
corrected for Galactic absorption of column density $N_{\rm H} =
1.6\times 10^{20}$~cm$^{-2}$ (Stark et al. 1992).

If a weakly-detected AGN does not have a $\ge 2 \sigma$ detection in
one or more of the 3 X-ray bands, we quote in Table 2 conservative $2
\sigma$ upper limits on the count rate and flux.  Upper limits were
calculated by adding any positive net source counts to 2 times the sky
dispersion.

\subsection{X-ray Coaddition}

For the five sources that are undetected and also a minimum distance
of 2 times their 95\% EER from the nearest cataloged X-ray source, we
determine limits on the X-ray flux by coaddition.  Individual source
counts are measured as described above, and are then summed to provide
the total number of counts from our stacked source.  We similarly
construct the coadded sky background distribution by adding the
individual sky backgrounds from each of the 10,000 trials.

We list in Table 3 the X-ray properties of the coadded source. Because
the detection significance for all 3 X-ray bands is $< 1 \sigma$, we
list conservative $2 \sigma$ and $5 \sigma$ upper limits on the count
rates and fluxes, calculated as described above.  We also list $2
\sigma$ and $5 \sigma$ upper limits on the flux for a "typical" one of
these five sources by assuming that an equal amount of flux is coming
from each of the sources that contributed to the coaddition.

\subsection{X-ray Upper Limits}

For all AGN that are not X-ray detected at full band to greater than
2$\sigma$, including those that are close to a known X-ray source, we
set conservative 5$\sigma$ upper limits on the X-ray flux by adding
any positive source flux to a 5$\sigma$ limit on the local X-ray
background, computed as described above. We use an aperture equal to
the 70\% EER to measure both the source and background counts.  These
5$\sigma$ X-ray upper limits are used throughout the rest of the
paper.

\subsection {Column Densities}

The column densities to the X-ray detected AGN are given in Table
4. To estimate the column density, we assume that AGN have an
intrinsic non-absorbed photon index of $\Gamma = 2$ (e.g., George et
al. 2000).  We do not include a Compton reflection component in our
intrinsic spectral model because of possible dependencies of its
strength on luminosity, radio-loudness, and optical Seyfert type
(e.g., Lawson \& Turner 1997, Reeves \& Turner 2000, Zdziarski et
al. 1995, Zdziarski et al. 2000).  These reflection components have
been shown to be important in fitting the peak of the X-ray background
(XRB; see Ueda et al. 2003). If a Compton reflection component is
added via the 'pexrav' model in {\sc XSPEC} (Magdziarz \& Zdziarski
1995), assuming a solid angle of 2$\pi$, a cutoff energy of $E_{\rm c}
= 500$~keV, an inclination angle of cos~$i$ = 0.5, and solar
abundance, the computed column densities decrease by an average of
20\%.

Excluding upper limits, the column densities of the X-ray-detected AGN
range from N$_{\rm H} = 1.0\times 10^{22}$~cm$^{-2}$ to N$_{\rm H} =
3.7\times 10^{23}$~cm$^{-2}$, and are therefore consistent with our
definition of obscured AGN.  One additional AGN for which we were
unable to estimate N$_{\rm H}$ (due to an observed photon index,
$\Gamma=2.1$, that is softer than our assumed photon index) is likely
to be relatively unobscured. All of the columns fall below the
Compton-thick limit (N$_{\rm H}>10^{24}$~cm$^{-2}$). If we correct for
the Compton-reflection component, 2 of the 11 X-ray detected AGN
(20\%) are likely to be unobscured.

We list in Table 2 intrinsic column densities for all weakly-detected
AGN.  For the two X-ray weakly-detected AGN with known flux ratios and
redshifts, we calculate column densities of N$_{\rm H} = 6.8\times
10^{22}$~cm$^{-2}$ and N$_{\rm H} = 1.6\times10^{23}$~cm$^{-2}$. We
place upper limits on the X-ray column densities of the remaining
X-ray weakly-detected AGN.  If we have no redshift estimate for a
given source, we calculate the column density at redshifts of $z=0.5$
and $z=2.5$, the approximate upper and lower redshift bounds of the
sources in our sample. All of the column densities fall below the
Compton-thick limit.

In Figure 4, we compare the column density distributions of the X-ray
detected and weakly-detected AGN.  The column densities for AGN with
known redshifts \textit{and} flux ratios are given by cross-hatched
histograms.  Clear histograms give the column densities for all AGN
with known redshifts, including those with upper limits on the flux
ratio, and therefore on the column density. There is significant
overlap between the 2 samples. The X-ray non-detected AGN are likely
to have higher column densities than the weakly-detected sample; if
so, there would be a stronger offset between the X-ray detected and
X-ray weakly- and non-detected samples.

\begin{figure}[t]
\epsscale{1.15}
\plotone{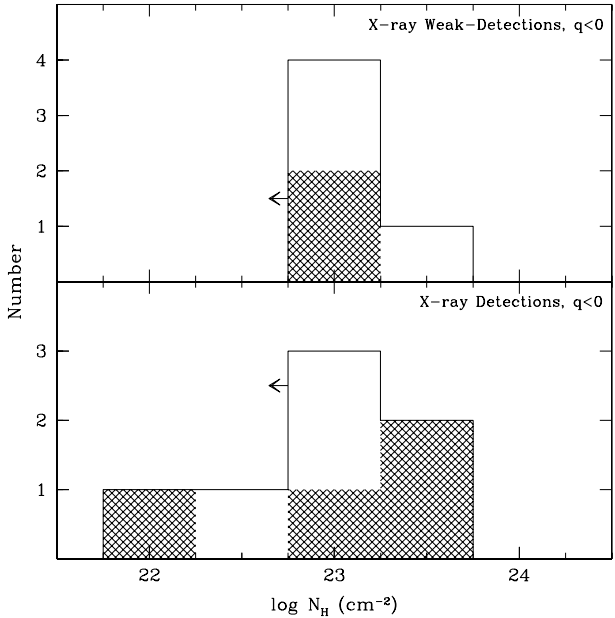}
\caption{Rest-frame column densities measured using the X-ray flux ratios
and redshifts of the X-ray weakly-detected and X-ray detected
radio-excess AGN samples. The clear histograms represent all sources,
including those for which we have only upper limits on $N_{\rm H}$.
The cross-hatched histograms represent only those sources with known
$N_{\rm H}$.}
\end{figure}

\subsection{Compton-thick AGN}

Approximately 50\% of Seyfert 2 galaxies in the local Universe are
Compton-thick (Risaliti et al. 1999); if we assume a Seyfert 2 to
Seyfert 1 ratio of 4:1, the Compton-thick fraction of all local
Seyferts is $\sim 40$\%. In the more distant Universe, XRB population
synthesis models predict that $\sim 30$\% of AGN with N$_{\rm H} <
10^{25}$~cm$^{-2}$ have column densities of N$_{\rm H} = 10^{24} -
10^{25}$~cm$^{-2}$ (Comastri et al. 2001). Ueda et al. (2003)
similarly find that current measurements of the XRB are consistent
with roughly equal numbers of Compton-thick AGN with N$_{\rm H} =
10^{24} - 10^{25}$~cm$^{-2}$ and obscured AGN with N$_{\rm H} =
10^{23} - 10^{24}$~cm$^{-2}$.  This ratio is slightly lower than that
in local Universe, where there are 1.6 times as many Seyfert 2's with
N$_{\rm H} > 10^{24}$~cm$^{-2}$ as Seyferts 2's with N$_{\rm H} =
10^{23} - 10^{24}$~cm$^{-2}$ (Risaliti et al. 1999).

We therefore expect that as many as $\sim 25-40$\% of the AGN in our
radio-selected sample could be Compton-thick.  Of the 27 radio-excess
AGN, we have placed constraints on the column densities of 13, and
have estimated constraints for a further 5 for which no redshift is
available. All sources have column densities that fall below the
Compton-thick limit. Of the remaining sources in the sample, 5 are
X-ray non-detected, and 4 are too close to a known X-ray source to
test for low-significance emission. If we assume that (like the
sources for which we could test for X-ray emission) $3/4$ of the
latter would be X-ray detected, 6 sources in our sample of 27 (22\%)
could be Compton-thick.  This estimate will improve as we increase our
sample by extending the current study to additional fields.


\section{Interpretation}

There is direct evidence, from the determination of X-ray column
densities discussed above, that at least several of the X-ray
weakly-detected AGN are highly obscured.  It appears, however, that
many of the X-ray detected AGN also have high column densities. To
better understand these two samples, and their relation to one
another, we discuss below their radio properties, their X-ray, radio,
and infrared luminosity distributions, and the ratios of their X-ray
to radio and X-ray to optical emission.

\subsection{Radio Extent and Spectral Shape}

We give in Table 1 the spatial extent of the radio emission, $\theta$,
as determined by Richards et al. (2000).  Of the 27 sample members, 1
has no listed extent, 16 are unresolved, and 10 are resolved. There is
no significant difference with X-ray flux in the degree of resolution,
at least in this small sample. The most highly resolved sources,
VLA~J123644+621133 and J123725+621128, also have high 1.4~GHz
luminosities and very low values of $q$.  It is likely that the large
radio spatial extent of these sources is due to powerful radio jets
oriented approximately in the plane of the sky.  If this is the case,
the AGN torus would be oriented edge-on, a condition that would lead
to high X-ray obscuration. Nonetheless, the absorption levels of
$N_{\rm H} < 2.0 \times 10^{22}$~cm$^{-2}$ and $N_{\rm H} < 2.0 \times
10^{23}$~cm$^{-2}$, respectively, are not exceptionally large.  It
follows that if the unresolved X-ray weakly- or non-detected AGN are
also viewed close to edge-on, as we would expect, they must not have
strong extended radio jets.  This is not surprising, however, as all
but 1 of these AGN (the exception being VLA~J123725+621128) have $q$
values that fall short of the radio-loud definition of Yun et
al. (2001).

We also list in Table 1 radio spectral slopes for the 17 radio-excess
AGN observed at 8.5~GHz by Richards et al.~(1998).  Following Richards
et al.~(2000), we define flat spectrum sources as those with $\alpha <
0.5$ and steep spectrum sources as those with $\alpha > 0.5$.  Flat
spectral slopes are expected for AGN in which synchrotron
self-absorption is important. Steep-spectrum emission with a slope of
$\alpha = 0.7$ is generally associated with optically thin synchrotron
radiation. Of the X-ray weakly- or non-detected AGN, 3 are flat
spectrum and 6 are steep spectrum. Five of the X-ray detected AGN are
flat spectrum; 3 are steep spectrum.  No obvious trend is seen,
although the X-ray detected AGN may generally have flatter spectral
slopes ($-0.28 < \alpha < 0.94$) than the X-ray non-detected AGN
($0.01 < \alpha < 1.62$).  This is consistent with a picture in which
our viewing angle to the X-ray detected AGN is aligned closer to the
jet axis, resulting in more core-dominated, and therefore flatter,
radio emission.

\subsection{Radio, 24~\micron, and X-ray Luminosities}

The 1.4~GHz radio powers, $P_{1.4~\rm GHz} = \nu F_{\nu}$, of the
radio-excess sample are shown in Figure 5.  Because radio emission is
not affected by obscuration, the radio power is a measure of the
intrinsic power of the sample. The X-ray weakly- or non-detected and
X-ray detected AGN span similar ranges of $P_{1.4~\rm GHz}$, although
the 2 most radio-luminous AGN have been missed in the X-ray. We do not
have redshift estimates for 4 of the X-ray weakly- or non-detected AGN
and 3 of the X-ray detected AGN; the true power distributions are
therefore likely to differ slightly from those presented here.

\begin{figure}[t]
\epsscale{1.15}
\plotone{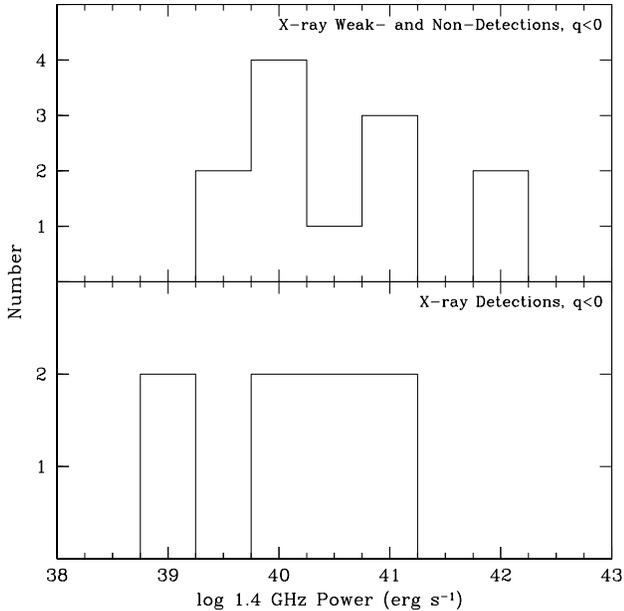}
\caption{1.4~GHz radio powers of the radio-excess AGN sample.}
\end{figure}

Within the limitations of the small number of 24$\mu$m detections, the
X-ray weakly- or non-detected and X-ray detected AGN also span similar
ranges of 24~\micron\ power.  Using the best-fit SEDs from our
redshift determinations, we can estimate $L_{\rm IR}$, the infrared
luminosity from $8-1000$~\micron.  We caution that these values are
only approximate, as the SEDS in the FIR are not well-constrained by
our data. We find that of the X-ray non- and weakly-detected AGN (with
redshift estimates), 1 is a ULIRG ($L_{\rm IR} > 10^{12}~\lsun$), 2
are LIRGS ($10^{11}~\lsun < L_{\rm IR} < 10^{12}~\lsun$), 8 have
luminosities of $10^{10}~\lsun < L_{\rm IR} < 10^{11}~\lsun$, and 1
has a luminosity that falls below $10^{10}~\lsun$. Two of the X-ray
detected AGN are ULIRGS, 1 is a LIRG, 3 have luminosities of
$10^{10}~\lsun < L_{\rm IR} < 10^{11}~\lsun$, and 2 have a luminosity
that falls below $10^{11}~\lsun$.

The 0.5-8~keV X-ray luminosities of our radio-excess sample are shown
in Figure 6. For all sources with column density estimates, we
calculate absorption-corrected X-ray luminosities, assuming an
intrinsic photon index of $\Gamma = 2$. The results are listed in
Tables 2 and 4. They are subject to relatively large errors, given the
uncertainties in the measurements of $\Gamma$ at low signal to noise
as well as possible intrinsic variations in the photon index. They are
nonetheless useful as rough estimates. Because there is no X-ray
K-correction for a spectrum with $\Gamma = 2$, rest-frame 0.5-8~keV
luminosities are equivalent to observed-frame, absorption-corrected
0.5-8~keV luminosities.  We overplot on Figure 6, as solid lines, the
rest-frame 0.5-8~keV X-ray luminosities of all sources with known
redshifts and column density estimates.  We see that of the AGN for
which we can estimate intrinsic luminosities, those with the highest
luminosities were detected in the X-ray.  Low-luminosity AGN (LLAGN),
however, were also detected in the X-ray; one such source,
VLA~J123652+621444, has an X-ray luminosity of $L_{0.5-8\ \rm{keV}} =
1.6\times 10^{41}$~erg~s$^{-1}$ and an optical through MIR SED that is
well fit by a LINER template.

\begin{figure}[t]
\epsscale{1.15}
\plotone{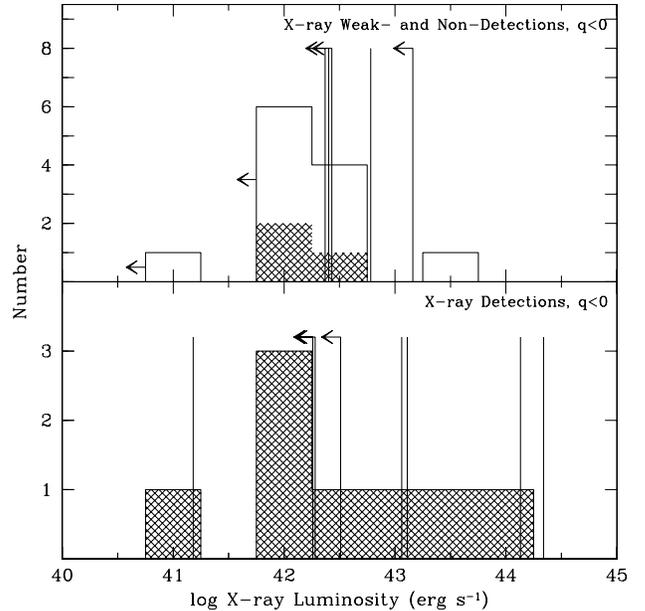}
\caption{Observed X-ray luminosity distributions of the radio-excess 
AGN sample.  The clear histograms represent all sources, including
those for which we have only upper limits on $L_{0.5-8\ \rm{keV}}$.
The cross-hatched histograms represent only those sources with known
$L_{0.5-8\ \rm{keV}}$. For those AGN for which we can estimate $N_{\rm
H}$, we overplot as solid lines the intrinsic 0.5-8~keV
absorption-corrected X-ray luminosities.}
\end{figure}

\subsection{Observed X-ray vs. Radio Emission}

The ratio of X-ray to radio emission provides another probe of the
nature of sources in our sample. We define \qprime\ = log (f$_{1.4~\rm
GHz}$/f$_{4 \rm keV}$) and compute the X-ray flux densities at an
energy of 4~keV (assuming a photon index of $\Gamma = 1.7$, equivalent
to a spectral index of $\alpha = 0.7$, for a bandpass
correction). Since a typical radio spectrum has a slope of $\alpha =
0.7$ ($S_{\rm \nu} \propto \nu^{-\alpha}$), we do not apply any
K-corrections to the ratio of radio to X-ray flux density. Thus, the
values we use are as close as possible to the observed radio vs. the
observed X-ray flux densities. Figure 7 shows the distribution of
\qprime\ for our X-ray detected and non-detected AGN samples, along
with all X-ray detections that meet our exposure cut.

\begin{figure}[t]
\epsscale{1.09}
\plotone{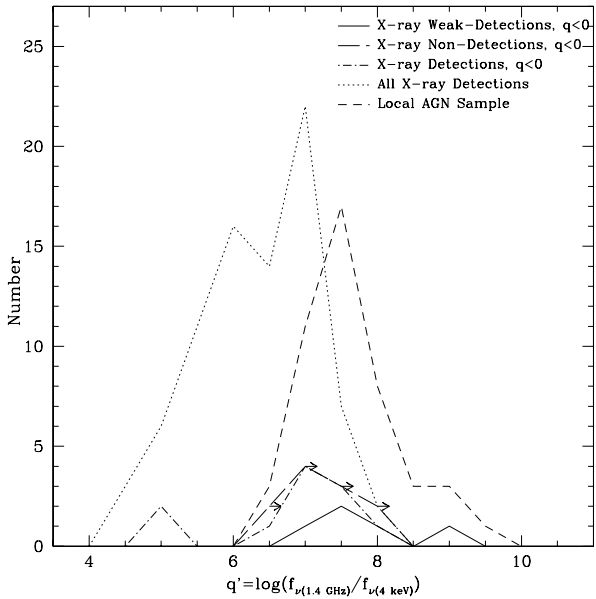}
\caption{Histogram of $q^\prime$. X-ray detections, weak detections,
and non-detections are samples as defined in this paper. The "local"
sample is from Sambruna et al. (1999), augmented by Gambill et
al. (2003). The sample labeled "all X-ray detections" represents the
X-ray sources in the CDFN that have radio counterparts in the catalog
of Richards (2000). For viewing clarity, we do not plot the full
histograms, but instead connect their values at the bin centers.  }
\end{figure}

To compare with a control sample, we have also plotted \qprime\ for
the X-ray detected (with {\it ASCA}) radio sample of Sambruna et
al. (1999). This study is of representative sources, not of a complete
well-defined sample, and contains 28 objects with sufficient data to
be plotted (of 38; the omitted sources are 3C99, 3C 275, 3C 295, 3C
330, 3C 368, 3C 411, 4C +41.17, 4C +55.16, 4C +74.26, PKS 0625-53, and
NGC 6251). All but four of the plotted sources are at z $\sim$ 0.3 or
less. All the objects for which IRAS 25~\micron\ data are available
have \qlt\ and are therefore AGN by the same criterion used in this
paper. Of the 28 sources, 16 are radio-intermediate (q $>$ -1.6) and
four more are only slightly more radio luminous (-1.6 $>$ q $>$
-2). They therefore are reasonable to compare with our predominantly
radio-intermediate sources.
 
We have also included sources from Gambill et al. (2003). This study
uses {\it Chandra} data for bright quasars over a large range of
redshift. The sample is mostly useful to augment the statistics for
radio quasars, which are not well represented in the Sambruna et
al. work.

Figure 7 shows that the overall behavior of \qprime\ is as one might
expect from selection biases: a sample of AGN selected in the X-ray
tends to be relatively brighter in X-rays compared with the radio than
does a sample of AGN selected in the radio (i.e. the control
sample). Because our use of \qprime\ normalizes both samples to radio
flux density, this behavior reflects an intrinsic difference in
radio-to-X-ray spectral energy distributions, and is not simply an
effect of low-luminosity radio sources without X-ray
counterparts. \textit{If the local radio-detected sample has analogs
at higher redshift, as expected, they would fall below the current
{\it Chandra} detection limit, yielding incomplete samples in the
X-ray.}

The \qprime\ distribution of the 5 X-ray weakly-detected sources is
clearly more consistent with that of the control sample than the full
{\it Chandra}-detected sample. The \qprime\ distributions also suggest
that some fraction of our X-ray non-detected AGN are more similar in
nature to these local AGN than they are to the full {\it
Chandra}-detected sample. In contrast, the radio-excess X-ray detected
AGN sample is more consistent with the overall {\it Chandra}-detected
sample than with the local AGN sample.  A two-sided Kolmogorov-Smirnov
(KS) test gives a probability of 4\% that the full {\it
Chandra}-detected and radio-excess X-ray detected AGN are drawn from
the same sample.  The corresponding probability for the {\it
Chandra}-detected and radio-excess X-ray weakly-detected AGN sample is
0.2\%.

The large values of \qprime\ for the control sample and the X-ray
weakly- or non-detected sample could be due to intrinsically lower
X-ray to radio luminosity, to heavy obscuration, or to some
combination of the two. We test the obscuration hypothesis by
comparing $R$, the ratio of compact to total radio emission, to
\qprime\ for the control sample. The compactness, log~$R$, is 
often used as an indication of the orientation of the radio source
relative to the observer. If the source is viewed down the jet axis,
log~$R = 0$, and the flux is dominated by beamed emission. As the
viewing angle moves away from the jet axis, log~$R$ becomes more
negative.

Figure 8 compares log~$R$ to \qprime\ for all of the control sample
except 3C28. We have excluded this object because the X-ray emission
from its position appears to be strongly dominated by emission by its
galaxy cluster, not the AGN (Hardcastle \& Worrall 1999). There is a
strong correlation in the sense that the more negative log~$R$, the
larger \qprime. That is, the more the radio jets are aligned away from
the observer, the weaker the X-rays are relative to the radio. This
behavior is consistent with a standard unification model where the
circumnuclear torus plays an increasing role in absorbing the X-rays
as the viewing angle of the central accretion disk approaches
edge-on. The parameter \qprime\ therefore appears to be correlated
with viewing angle, and consequently with obscuration, in local
samples. We note that the 2 highly radio-resolved sources discussed in
\S6.1 also have the highest values of \qprime, a result that is
consistent with the above interpretation.  The relatively high limits
on \qprime\ for the X-ray non-detected AGN therefore suggest that the
beamed radio emission of these AGN is aligned away from us, increasing
their likelihood of being heavily obscured.

\begin{figure}[t]
\epsscale{1.125}
\plotone{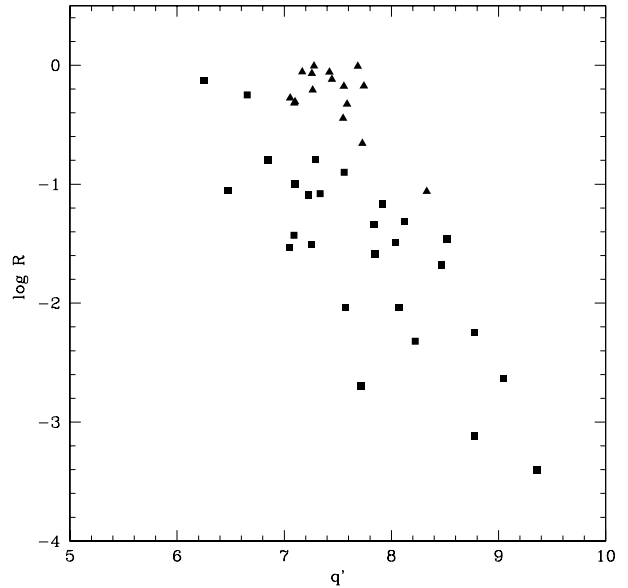}
\caption{Relationship between log $R$, the ratio of compact to total 
radio emission, and \qprime\ for the local AGN samples of Sambruna et
al. (1999; squares) and Gambill et al. (2003; triangles). }
\end{figure}

\subsection{Intrinsic X-ray vs. Radio Emission}

We plot in Figure 9 the intrinsic X-ray to radio emission ratios,
\qprimecorr, for the X-ray weakly-detected and X-ray detected samples.
We calculate \qprimecorr\ by correcting the observed 0.5-8~keV X-ray
flux for the estimated column densities. We assume a photon index of
$\Gamma = 2$ when applying this correction. Because the radio spectral
slopes are known for the majority of the sources for which we have
column density estimates, we apply a radio K-correction, such that
\qprimecorr\ $=$ log (f$_{1.4~\rm GHz}$/f$_{4 \rm
keV, corr}$) + $(\alpha_{\rm radio} - 1)$log$(1+z)$. We list the
values of \qprimecorr\ in Tables 2 and 4. For sources without known
radio spectral indices, we assume $\alpha = 0.7$.

\begin{figure}[t]
\epsscale{1.1}
\plotone{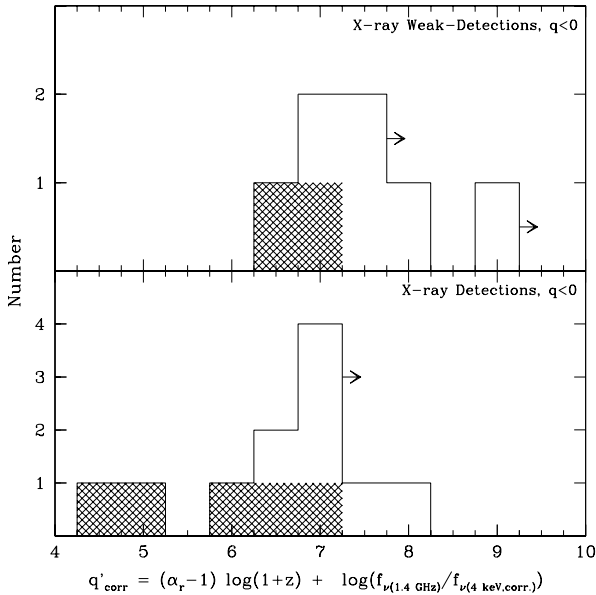}
\caption{Histogram of the absorption- and K-corrected $q^\prime$. 
The clear histograms represent all sources, including those for which
we have only lower limits on $q^{\prime}_{\rm{corr}}$.  The
cross-hatched histograms represent only those sources with known
$q^{\prime}_{\rm{corr}}$. }
\end{figure}

The X-ray detected sample has, on average, a lower \qprimecorr\ than
the X-ray non-detected sample, although there is a significant amount
of overlap between the 2 samples and many of the sources have only
lower limits on \qprimecorr.  This suggests, however, that some
fraction of the X-ray detected sample may have intrinsically higher
X-ray luminosities (with respect to their radio luminosities) than do
sources that were not detected in the X-ray.  Such a conclusion is in
agreement with the absorption-corrected X-ray luminosities, which are
generally higher for the X-ray--detected AGN.  A sample of
intrinsically X-ray weak AGN with hard X-ray spectra was studied by
Risaliti et al. (2003), who suggest that such objects could contribute
significantly to the hard component of the X-ray
background. Unfortunately, because we do not have column density
estimates for the X-ray non-detected sources, it is unclear whether
this trend will hold for the entire sample.

It is also possible that a selection effect is responsible for the
apparently lower intrinsic X-ray luminosities of the X-ray
weakly-detected sample.  Because the X-ray detected and X-ray
weakly-detected samples span similar ranges of column densities, it is
likely that the AGN that were previously X-ray detected are those that
are intrinsically the most luminous.

\subsection{X-ray vs. Optical Emission}

The optical to X-ray emission ratio is a powerful diagnostic for
distinguishing AGN from star-forming galaxies (e.g. Maccacaro et
al. 1988).  AGN typically lie within log $f_{\rm X}/f_{\rm R} = \pm
1$; at the highest fluxes probed by the CDFN, the mean AGN flux ratio
is consistent with log $f_{\rm X}/f_{\rm R} = 0$, whereas at lower
fluxes, the mean ratio rises slightly due to increased optical
obscuration (Barger et al.~2003). Starbursts and lower-luminosity AGN
populate the transition region ($-2 <$ log$f_{\rm X}/f_{\rm R} < -1$),
whereas normal star-forming galaxies, starbursts, and LLAGN typically
have X-ray to optical flux ratios of log~$f_{\rm X}/f_{\rm R} < -2$
(see Barger et al. 2003, Hornschemeier et al. 2003)

Figure 10 shows the observed R-band magnitudes and hard (2-8~keV)
X-ray fluxes of our radio-excess AGN sample. We place a lower limit of
R=25.0~mag on sources with no R-band counterpart. For sources with
multiple GOODS counterparts, but only a single ground-based R-band
counterpart, we treat the R-band magnitude as a lower limit.

\begin{figure}[t]
\epsscale{1.12}
\plotone{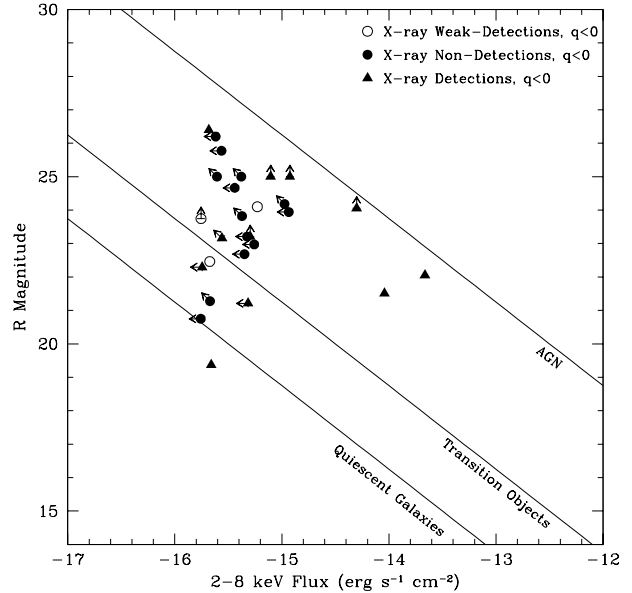}
\caption{Relationship between the observed R magnitude and 2-8~keV (hard band) 
X-ray flux.  X-ray detected sources are given by triangles, and X-ray
weakly- and non-detected sources by circles. The lines represent the
regions populated by AGN ($f_{\rm X}/f_{\rm R} < |1|$), AGN and
starbursts ($-2 <$ log$f_{\rm X}/f_{\rm R} < -1$), and quiescent
galaxies, starbursts, and LLAGN (log $f_{\rm X}/f_{\rm R} < -2$) (see
Barger et al. 2003, Hornschmeier et al. 2003).}
\end{figure}

Of the 11 hard X-ray-detected radio-excess AGN, all but 1 lie within
the AGN or transition regions, as expected.  The source that lies in
the quiescent galaxy region is the LLAGN, VLA~J123652+621444,
discussed in \S6.2. The current X-ray and optical limits also place
the X-ray weakly- and non-detected AGN primarily in the AGN region,
with 3 extending into the transition region.  Higher sensitivity X-ray
observations are needed to determine the true locations of the sample.
Given the current limits, however, all 16 sources are consistent with
the predictions for normal AGN.


\section{Conclusions} 

We define a sample of 27 radio-excess AGN in the CDFN by selecting
galaxies that do not obey the radio/infrared relation for star-forming
galaxies and radio-quiet AGN. Sixteen (60\%) of these AGN are
previously undetected at X-ray exposures of $>$ 1~Ms, although we have
subsequently detected X-ray signals from 7 more at significance levels
of 2.5 to 5.2 $\sigma$. In this paper, we examine the characteristics
of these 16 X-ray weak AGN, and their relationship to the 11 AGN
previously detected in the X-ray.  We find:

\begin{itemize}
\item
The values and limits we place on N$_H$ for the X-ray weak AGN are all
consistent with their being obscured, but not Compton-thick (i.e.,
$10^{22}$ cm$^{-2} <$ N$_H$ $< 10^{24}$ cm$^{-2}$).  Of the 9 AGN that
are either X-ray non-detected or are too close to a known X-ray source
to search for weak X-ray emission, we estimate that 6 could be Compton
thick.  This corresponds to 22\% of our sample and is consistent with
predictions from the X-ray background.

\item
The column densities to the X-ray-detected members of our sample are
similar to those of the X-ray weak AGN, although they extend to lower
absorbing column.  All but 1 (or 2 if a Compton reflection component
is considered) fall in the range of obscured but not Compton-thick
AGN.  This corresponds to 80\% of the sample, and is consistent with
predictions that X-ray obscured AGN outnumber non-obscured AGN by a
factor of 4:1.  If we assume that all of the X-ray non-detected AGN
are obscured, however, the overall fraction of obscured AGN rises to
93\%.

\item
The ratio of radio to X-ray emission indicates that orientation of the
circumnuclear torus, and therefore obscuration, is at least partly
responsible for attenuating the X-ray fluxes, as we would expect. To
support this scenario, we show that a local sample of radio AGN has a
ratio of radio to X-ray emission that shows a dependence on viewing
angle consistent with the effects of X-ray absorption by a
circumnuclear torus.

\item
Indicators such as the ratio of optical to X-ray emission, the
intrinsic absorption-corrected X-ray luminosities, and the radio
luminosities all support the argument that the weak X-ray sources are
similar in intrinsic properties to the AGN selected in a similar way
but with stronger X-ray fluxes. This similarity would be strengthened
by the usual tendency that any sample selected in a given spectral
band (e.g., X-ray or radio) will tend to include preferentially the
sources of a given population that are relatively brighter in that
band.
\end{itemize}

In summary, $\sim$ 60\% of radio-excess AGN have been missed in the
CDFN survey at X-ray exposures $>$ 1 Ms. At least $1/2$ of the
radio-excess AGN have high, but not extreme column densities (
$10^{22}$~cm$^{-2} <$ N$_H$ $< 10^{24}$~cm$^{-2}$), and X-ray, radio,
and infrared luminosities similar to those of normal radio-excess
X-ray detected AGN.  Six (22\%) of the sample may be Compton thick
(N$_{\rm H} > 10^{24}$~cm$^{-2}$).  These results are consistent with
proposals that obscured AGN outnumber unobscured ones by $\sim$~4:1,
although the proportion of potentially Compton-thick objects is at the
low end of the predictions.


\acknowledgments 

This work was supported by an NSF Graduate Research Fellowship and by
NASA through contract 960785 issued by JPL/California Institute of
Technology. We thank Franz Bauer for an exceptionally detailed and
helpful referee report.


\clearpage


\clearpage
\begin{landscape}
\begin{deluxetable}{lllccccrrrrlrl}
\pagestyle{empty}
\tablenum{1}
\tabletypesize{\tiny}
\tablewidth{0pt}
\tablecaption {Radio, 24 \micron, and X-ray Properties}
\tablehead{
\colhead{$\alpha_{\rm 2000}$}                  &
\colhead{$\delta_{\rm 2000}$}                  &
\colhead{z}                                    &
\colhead{1.4 GHz}                              &
\colhead{log 1.4 GHz}                          &
\colhead{24 $\micron$}                         &
\colhead{log 24 $\micron$}                     &
\colhead{X-ray\tablenotemark{a}}               &
\colhead{log X-ray}                            &
\colhead{$q$}                                  &
\colhead{$q^{\prime}$}                         &
\colhead{$\alpha$}                             &
\colhead{$\theta$}                             &
\colhead{z Ref.}                               \\
\colhead{}                                     &
\colhead{}                                     &
\colhead{}                                     &
\colhead{flux}                                 &
\colhead{Power}                                &
\colhead{flux}                                 &
\colhead{Power}                                &
\colhead{flux}                                 &
\colhead{Luminosity}                           &
\colhead{}                                     &
\colhead{}                                     &
\colhead{}                                     &
\colhead{(\arcsec)}                            &
\colhead{}                                     \\
\colhead{}                                     &
\colhead{}                                     &
\colhead{}                                     &
\colhead{($\mu$Jy)}                            &
\colhead{(erg s$^{-1}$)}                       &
\colhead{($\mu$Jy)}                            &
\colhead{(erg s$^{-1}$)}                       &
\colhead{(erg cm$^{2}$ s$^{-1}$)}              &
\colhead{(erg s$^{-1}$)}                       &
\colhead{}                                     &
\colhead{}                                     &
\colhead{}                                     &
\colhead{}                                     &
\colhead{}                                     
}
\startdata 
\cutinhead{X-ray Weakly-Detected and Non-Detected AGN}
12 35 47.96 &  62 15 29.2 & \nodata   & 118    & \nodata & 40.8    & \nodata   &  $<$2.28$\times 10^{-16}$    & \nodata   &    -0.46  & $>$7.1  & \nodata        &  $<$2.1    & \nodata   \\
12 36 10.57 &  62 16 51.6 & 1.5       & 139    &   40.4  & $<$80   &  $<$44.1  &  $<$1.57$\times 10^{-16}$    & $<$42.3   & $<$-0.24  & $>$7.3  & \nodata        &  2.7       & 7         \\
12 36 17.57 &  62 15 40.7 & \nodata   & 200    & \nodata & $<$80   & \nodata   &     1.33$\times 10^{-16}$    & \nodata   & $<$-0.40  &    7.5  & $>$0.55        &  $<$2.0    & \nodata   \\
12 36 20.28 &  62 08 44.1 & 1.02      & 123    &   40.0  & $<$80   &  $<$43.7  &     2.56$\times 10^{-16}$    &    42.1   & $<$-0.19  &    7.0  & 0.01$\pm$0.07  &  $<$2.1    & 1,2       \\
12 36 23.54 &  62 16 42.7 & \nodata   & 481    & \nodata & $<$80   & \nodata   &     9.64$\times 10^{-17}$    & \nodata   & $<$-0.78  &    8.0  & 0.63$\pm$0.07  &  $<$1.9    &\nodata    \\
12 36 40.15 &  62 20 37.4 & 0.8       & 1840   &   40.9  & $<$80   &  $<$43.5  &  $<$3.02$\times 10^{-16}$    & $<$41.9   & $<$-1.36  & $>$8.1  & \nodata        &  $<$1.9    & 7         \\
12 36 40.74 &  62 10 10.6 & 1.3       & 86.8   &   40.1  & $<$80   &  $<$44.0  &  $<$1.38$\times 10^{-16}$    & $<$42.1   & $<$-0.04  & $>$7.1  & 0.44$\pm$0.15  &  3.7       & 7         \\
12 36 46.34 &  62 16 29.6 & 0.50      & 393    &   39.7  & $<$80   &  $<$43.0  &  $<$1.10$\times 10^{-16}$    & $<$41.0   & $<$-0.69  & $>$7.9  & $>$1.62        &  2.7       & 1,2       \\
12 37 00.26 &  62 09 9.8  & 1.8       & 324    &   41.0  & 203     &     44.7  &  $<$1.57$\times 10^{-16}$    & $<$42.5   &    -0.20  & $>$7.7  & 0.89$\pm$0.12  &  $<$2.0    & 7         \\
12 37 08.78 &  62 22 01.9  & \nodata   & 170    & \nodata & $<$80   & \nodata   &  $<$6.97$\times 10^{-16}$    & \nodata   & $<$-0.33  & $>$6.7  & \nodata        &  $<$2.1    & \nodata   \\
12 37 14.96 &  62 08 23.5 & 0.9       & 1380   &   40.9  & 255     &     44.1  &     5.97$\times 10^{-16}$    &    42.4   &    -0.73  &    7.7  & 0.15$\pm$0.08  &  $<$1.9    & 7         \\
12 37 17.53 &  62 08 27.7 & 0.9       & 126    &   39.9  & 106     &     43.7  &  $<$3.00$\times 10^{-16}$    & $<$42.1   &    -0.08  & $>$7.0  & $>$0.77        &  2.1       & 7         \\
12 37 25.73 &  62 11 28.5  & 1.2       & 5960   &   41.8  & $<$80   &  $<$43.9  &     1.50$\times 10^{-16}$    &    42.1   & $<$-1.87  &    9.0  & 1.35$\pm$0.06  &  5.5       & 7         \\
12 37 45.73 &  62 14 56.3 & 0.9       & 63.6   &   39.6  & 59.4    &     43.5  &  $<$2.64$\times 10^{-16}$    & $<$42.0   &    -0.03  & $>$6.7  & \nodata        &  \nodata   & 7         \\
12 37 46.60 &  62 17 38.4 & 2.4       & 998    &   41.8  & 192     &     45.0  &  $<$4.22$\times 10^{-16}$    & $<$43.3   &    -0.72  & $>$7.7  & \nodata        &  $<$1.9    & 7         \\
12 37 51.21 &  62 19 19.1 & 1.0       & 223    &   40.2  & $<$80   &  $<$43.7  &  $<$6.25$\times 10^{-16}$    & $<$42.5   & $<$-0.45  & $>$6.9  & \nodata        &  $<$2.0    & 7         \\
\cutinhead{X-ray Detected AGN}
12 35 56.05  & 62 15 55.6 &    0.44    &  89.7   &   38.9   &   26.8   &    42.3  &  1.31$\times 10^{-14}$ &       42.9 &     -0.52 &    5.2  & \nodata         &    2.5 & 3       \\
12 36 40.6   & 62 18 33    &    1.14    &   324   &   40.5   &  $<$205  & $<$44.2  &  2.13$\times 10^{-16}$ &       42.2 &  $<$-0.20 &    7.5  & $>0.8$          & $<$2.0 & 1       \\
12 36 42.09   & 62 13 31.4 & \nodata    &   467   &\nodata   &    197   & \nodata  &  2.76$\times 10^{-16}$ &    \nodata &     -0.37 &    7.6  & $0.94\pm0.06$   &    2.2 & 4,5     \\
12 36 44.39  & 62 11 33.1  &    1.01    &  1290   &   41.0   &  $<$80   & $<$43.7  &  2.44$\times 10^{-16}$ &       42.1 &  $<$-1.21 &    8.1  & $0.3\pm0.05$    &    12  & 1,2     \\
12 36 46.34   & 62 14 4.7  &    0.96    &   179   &   40.1   &    162   &    44.0  &  2.46$\times 10^{-14}$ &       44.0 &     -0.04 &    5.2  & $-0.04\pm0.06$  & $<$2.0 & 1,2     \\
12 36 52.92  & 62 14 44.0 &    0.32    &   168   &   38.9   &  $<$80   & $<$42.5  &  4.63$\times 10^{-16}$ &       41.2 &  $<$-0.32 &    6.9  & $-0.12\pm0.07$  &    2.5 & 1,2     \\
12 36 59.15  & 62 18 32.8 & \nodata    &   506   &\nodata   &    329   & \nodata  &  8.76$\times 10^{-16}$ &    \nodata &     -0.19 &    7.1  & $0.26\pm0.07$   & $<$2.0 & \nodata \\
12 37 1.10   & 62 21 9.6   &     0.8    &   304   &$<$40.1   &  $<$289  & $<$44.0  &  3.22$\times 10^{-16}$ &       42.0 &  $<$-0.02 &    7.3  & \nodata         & $<$2.0 & 1,2     \\
12 37 9.94    & 62 22 58.9 &    1.23    &   708   &   40.9   &    370   &    44.6  &  5.21$\times 10^{-15}$ &       43.6 &     -0.28 &    6.5  & \nodata         & $<$1.9 & 6       \\
12 37 13.87   & 62 18 26.5 & \nodata    &   595   &\nodata   &  $<$80   & \nodata  &  1.69$\times 10^{-15}$ &    \nodata &  $<$-0.87 &    6.9  & $>0.92$         & $<$1.9 & \nodata \\
12 37 21.25  & 62 11 30.0 &    1.3     &   382   &   40.7   &    250   &    44.5  &  5.09$\times 10^{-16}$ &       42.7 &     -0.18 &    7.2  & $-0.28\pm0.06$  &    0.9 & 7       \\
\enddata
\tablenotetext{a}{5$\sigma$ upper limits are given for all AGN that were not weakly-detected in the X-ray (see \S5.3)}
\tablerefs{(1) Cowie et al. 2004 (Hawaii); (2) Wirth et al. 2004 (TKRS); (3) Barger et al. 2002; (4) Waddington et al. 1999; (5) Barger et al. 2000; (6) Barger et al. 2003 (photometric); (7) our photometric estimate}

\end{deluxetable}
\clearpage
\end{landscape}


\clearpage
\begin{landscape}
\begin{deluxetable}{llclccrrrrcccc}
\tablenum{2}
\tabletypesize{\tiny}
\tablewidth{0pt}
\tablecaption {X-ray Properties of X-ray Weakly-Detected AGN}
\tablehead{
\colhead{Source}                 &
\colhead{z}                      &
\colhead{Nearest}                & 
\colhead{X-ray}                  &
\colhead{EER}                    &
\colhead{Detection }             &
\colhead{Aperture}               &
\colhead{$H/S$}                  &
\colhead{$\Gamma$}               &
\colhead{Flux}                   &
\colhead{Hard/Soft}              &
\colhead{N$_{\rm H}$}            &
\colhead{Abs. Corrected }        &
\colhead{q$^{\prime}_{\rm {corr}}$}\\
\colhead{}                       &
\colhead{}                       &
\colhead{X-ray Source\tablenotemark{a}}  &
\colhead{Band}                   &
\colhead{(\%)}                   &
\colhead{$\sigma$}               &
\colhead{Corrected}              &
\colhead{}                       &
\colhead{}                       &
\colhead{(erg s$^{-1}$ cm$^{-2}$)} &
\colhead{Flux }                  &
\colhead{(cm$^{-2}$)}            &
\colhead{log L$_{0.5-8\ \rm{keV}}$} &
\colhead{}                       \\
\colhead{}                       &
\colhead{}                       &
\colhead{}                       &
\colhead{}                       &
\colhead{}                       &
\colhead{}                       &
\colhead{Count Rate (s${^{-1}}$)} &
\colhead{}                       &
\colhead{}                       &
\colhead{}                       &
\colhead{Ratio}                  &
\colhead{}                       &
\colhead{(erg s$^{-1}$)}         &
\colhead{}                       
}
\setlength{\tabcolsep}{0.02in}
\startdata         
VLA J123617+621540 &\nodata & 20.4(4.3) & Full  & 70  &  3.1  &    7.06$\times 10^{-06}$  & 1.5    & 0.47      &    1.33$\times 10^{-16}$  & 8.6      & (5.0$\times 10^{22}$-5.6$\times 10^{23}$)        & (41.4-43.2)         & 7.2-7.0  \\ 	    
                   &        &           & Hard  & 80  &  2.5  &    6.14$\times 10^{-06}$  &        &           &    1.76$\times 10^{-16}$  &          &                                                  & (41.3-43.1)         &          \\ 	              
                   &        &           & Soft  & 60  &  3.9  &    4.15$\times 10^{-06}$  &        &           &    1.99$\times 10^{-17}$  &          &                                                  & (41.3-43.1)         &          \\ 	              
VLA J123620+620844 & 1.02   & 29.5(5.6) & Full  & 60  &  3.6  &    1.60$\times 10^{-05}$  & 1.0    & 0.80      &    2.56$\times 10^{-16}$  & 5.4      & 6.8$\times 10^{22}$                              &    42.4             & 6.5      \\ 	    
                   &        &           & Hard  & 60  &  2.1  &    8.02$\times 10^{-06}$  &        &           &    2.12$\times 10^{-16}$  &          &                                                  &    42.1             &          \\ 	              
                   &        &           & Soft  & 60  &  3.5  &    7.79$\times 10^{-06}$  &        &           &    3.78$\times 10^{-17}$  &          &                                                  &    42.1             &          \\ 	              
VLA J123623+621642 &\nodata & 14.3(3.1) & Full  & 60  &  3.6  &    8.27$\times 10^{-06}$  & $<$0.7 & $>$1.09   &    9.64$\times 10^{-17}$  & $<$3.3   & ($<$1.8$\times 10^{22}$-$<$1.8$\times 10^{23}$)  &  ($<$41.2-$<$42.9)  & $>$7.9-$>$7.7  \\  
                   &        &           & Hard  & 70  &  0.9  & $<$4.56$\times 10^{-06}$  &        &           & $<$1.05$\times 10^{-16}$  &          &                                                  &  ($<$41.0-$<$42.8)  &          \\ 	              
                   &        &           & Soft  & 60  &  5.2  &    6.12$\times 10^{-06}$  &        &           &    3.07$\times 10^{-17}$  &          &                                                  &  ($<$41.0-$<$42.8)  &          \\ 	              
VLA J123700+620909 & 1.8    & 31.2(5.9) & Full  & 60  &  0.7  & $<$5.36$\times 10^{-06}$  & $<$1.3 & $>$0.60   & $<$6.25$\times 10^{-17}$  & $<$5.7   & $<$1.9$\times 10^{23}$                           & $<$42.4             & $>$7.8   \\ 	    
                   &        &           & Hard  & 80  & -0.7  & $<$5.57$\times 10^{-06}$  &        &           & $<$1.28$\times 10^{-16}$  &          &                                                  & $<$42.5             &          \\ 	              
                   &        &           & Soft  & 70  &  2.5  &    4.33$\times 10^{-06}$  &        &           &    2.17$\times 10^{-17}$  &          &                                                  & $<$42.5             &          \\ 	              
VLA J123714+620823 & 0.9    & 27.2(3.7) & Full  & 60  &  5.0  &    2.46$\times 10^{-05}$  & 2.7    & -0.06     &    5.97$\times 10^{-16}$  & 17.7     & 1.6$\times 10^{23}$                              &    42.8             & 7.1      \\ 	    
                   &        &           & Hard  & 60  &  4.2  &    1.82$\times 10^{-05}$  &        &           &    5.91$\times 10^{-16}$  &          &                                                  &    42.5             &          \\ 	              
                   &        &           & Soft  & 60  &  2.8  &    6.83$\times 10^{-06}$  &        &           &    3.26$\times 10^{-17}$  &          &                                                  &    42.5             &          \\ 	              
VLA J123725+621128 & 1.2    & 31.1(5.4) & Full  & 80  &  2.9  &    1.29$\times 10^{-05}$  & $<$1.4 & $>$0.53   &    1.50$\times 10^{-16}$  & $<$6.1   & $<$1.0$\times 10^{23}$                           & $<$42.4             & $>$8.8   \\ 	    
                   &        &           & Hard  & 70  &  1.5  & $<$8.52$\times 10^{-06}$  &        &           & $<$1.96$\times 10^{-16}$  &          &                                                  & $<$42.2             &          \\ 	              
                   &        &           & Soft  & 80  &  2.8  &    6.15$\times 10^{-06}$  &        &           &    3.08$\times 10^{-17}$  &          &                                                  & $<$42.2             &          \\ 	              
VLA J123746+621738 & 2.4    & 44.1(5.2) & Full  & 60  &  1.8  & $<$1.69$\times 10^{-05}$  & $<$0.7 & $>$1.12   & $<$1.97$\times 10^{-16}$  & $<$3.2   & $<$1.6$\times 10^{23}$                           & $<$43.2             & $>$7.7   \\ 	    
                   &        &           & Hard  & 80  &  0.0  & $<$1.18$\times 10^{-05}$  &        &           & $<$2.72$\times 10^{-16}$  &          &                                                  & $<$43.1             &          \\ 	              
                   &        &           & Soft  & 70  &  4.1  &    1.65$\times 10^{-05}$  &        &           &    8.27$\times 10^{-17}$  &          &                                                  & $<$43.1             &          \\ 	              
\enddata
\tablenotetext{a}{Nearest X-ray source in arcseconds and ratio of this distance to the hard-band 95\% encircled energy aperture radius}
\end{deluxetable}

\clearpage
\end{landscape}

\clearpage
\begin{landscape}
\begin{deluxetable}{lrrrrr}
\tablenum{3}
\tabletypesize{\tiny}
\tablewidth{0pt}
\tablecaption {Limits on the X-ray flux of 5 X-ray Non-Detected AGN}
\tablehead{
\colhead{Band}                   &
\colhead{2$ \sigma$}             &
\colhead{5$ \sigma$}             &
\colhead{2$ \sigma$ Individ.}    &
\colhead{5$ \sigma$ Individ.}    \\
\colhead{}                       &
\colhead{Flux Limit}             &
\colhead{Flux Limit}             &
\colhead{Flux Limit}             &
\colhead{Flux Limit}            \\
\colhead{}                       &
\colhead{(erg s$^{-1}$ cm$^{-2}$)} &
\colhead{(erg s$^{-1}$ cm$^{-2}$)} &                      
\colhead{(erg s$^{-1}$ cm$^{-2}$)} &
\colhead{(erg s$^{-1}$ cm$^{-2}$)}                       
}
\startdata  
Full  & $<$4.11$\times 10^{-16}$  & $<$8.42$\times 10^{-16}$  & $<$8.23$\times 10^{-17}$  & $<$1.68$\times 10^{-16}$     \\ 
Hard  & $<$6.50$\times 10^{-16}$  & $<$1.41$\times 10^{-15}$  & $<$1.30$\times 10^{-16}$  & $<$2.81$\times 10^{-16}$     \\ 
Soft  & $<$8.32$\times 10^{-17}$  & $<$1.79$\times 10^{-16}$  & $<$1.66$\times 10^{-17}$  & $<$3.59$\times 10^{-17}$     \\ 
\enddata
\end{deluxetable}

\clearpage
\end{landscape}

\clearpage
\begin{landscape}
\begin{deluxetable}{llllrcccc}
\tablenum{4}
\tabletypesize{\tiny}
\tablewidth{0pt}
\tablecaption {X-ray Properties of X-ray Detected AGN}
\tablehead{
\colhead{Source}                 &
\colhead{z}                      &
\colhead{X-ray}                  &
\colhead{$\Gamma$\tablenotemark{a}} &
\colhead{Flux}                   &
\colhead{Hard/Soft}              &
\colhead{N$_{\rm H}$}            &
\colhead{Abs. Corrected}         &
\colhead{q$^{\prime}_{\rm {corr}}$} \\
\colhead{}                       &
\colhead{}                       &
\colhead{Band}                   &
\colhead{}                       &
\colhead{(erg s$^{-1}$ cm$^{-2}$)} &
\colhead{Flux}                   &
\colhead{(cm$^{-2}$)}            &
\colhead{log L$_{\rm x}$}        &
\colhead{}                      \\
\colhead{}                       &
\colhead{}                       &
\colhead{}                       &
\colhead{}                       &
\colhead{}                       &
\colhead{Ratio}                  &
\colhead{}                       &
\colhead{(erg s$^{-1}$)}         &
\colhead{}                       
}

\startdata         
VLA J123556+621555   &    0.44   & Full  & 1.37     &    1.31$\times 10^{-14}$  &    2.4   &    1.0$\times 10^{22}$                     & 43.1        & 5.0       \\  
                     &           & Hard  &          &    9.10$\times 10^{-15}$  &          &                                            & 42.8        &           \\           
                     &           & Soft  &          &    3.62$\times 10^{-15}$  &          &                                            & 42.8        &           \\           
VLA J123640+621833   &    1.14   & Full  & >0.26    &    2.13$\times 10^{-16}$  & $<$8.3   &   <1.3$\times 10^{23}$                     & $<$42.5     & $>$7.2    \\  
                     &           & Hard  &          & $<$2.76$\times 10^{-16}$  &          &                                            & $<$42.4     &           \\           
                     &           & Soft  &          &    3.20$\times 10^{-17}$  &          &                                            & $<$42.4     &           \\           
VLA J123642+621331   & \nodata   & Full  & 1.35     &    2.76$\times 10^{-16}$  &    2.5   & (1.2$\times 10^{22}$-1.1$\times 10^{23}$)  & (41.6-43.3) & 7.5- 7.4  \\  
                     &           & Hard  &          &    2.08$\times 10^{-16}$  &          &                                            & (41.3-43.0) &           \\           
                     &           & Soft  &          &    8.03$\times 10^{-17}$  &          &                                            & (41.3-43.0) &           \\           
VLA J123644+621133   &    1.01   & Full  & >1.43    &    2.44$\times 10^{-16}$  & $<$2.2   & $<$2.0$\times 10^{22}$                     & $<$42.3     & $>$7.8    \\  
                     &           & Hard  &          & $<$1.80$\times 10^{-16}$  &          &                                            & $<$42.0     &           \\           
                     &           & Soft  &          &    7.74$\times 10^{-17}$  &          &                                            & $<$42.0     &           \\           
VLA J123646+621404   &    0.96   & Full  & 0.67     &    2.46$\times 10^{-14}$  &    6.4   &    7.6$\times 10^{22}$                     & 44.3        & 4.7       \\  
                     &           & Hard  &          &    2.17$\times 10^{-14}$  &          &                                            & 44.1        &           \\           
                     &           & Soft  &          &    3.26$\times 10^{-15}$  &          &                                            & 44.0        &           \\           
VLA J123652+621444   &    0.32   & Full  & 2.10     &    4.63$\times 10^{-16}$  &    0.9   &   \nodata                                  & 41.2        & 6.8       \\  
                     &           & Hard  &          &    2.19$\times 10^{-16}$  &          &                                            & 40.8        &           \\           
                     &           & Soft  &          &    2.44$\times 10^{-16}$  &          &                                            & 40.9        &           \\           
VLA J123659+621832   & \nodata   & Full  & 0.67     &    8.76$\times 10^{-16}$  &    6.5   & (3.9$\times 10^{22}$-4.3$\times 10^{23}$)  & (42.2-44.0) & 6.8-6.4   \\  
                     &           & Hard  &          &    7.86$\times 10^{-16}$  &          &                                            & (41.9-43.7) &           \\           
                     &           & Soft  &          &    1.18$\times 10^{-16}$  &          &                                            & (41.9-43.7) &           \\           
VLA J123701+622109   &     0.8   & Full  & >0.40    &    3.22$\times 10^{-16}$  & $<$7.1   & $<$6.8$\times 10^{22}$                     & $<$42.3     & $>$7.0    \\  
                     &           & Hard  &          & $<$4.83$\times 10^{-16}$  &          &                                            & $<$42.2     &           \\           
                     &           & Soft  &          &    6.56$\times 10^{-17}$  &          &                                            & $<$42.2     &           \\           
VLA J123709+622258   &    1.23   & Full  & -0.58    &    5.21$\times 10^{-15}$  &   36.2   &    3.7$\times 10^{23}$                     & 44.1        & 6.0       \\  
                     &           & Hard  &          &    4.99$\times 10^{-15}$  &          &                                            & 43.8        &           \\           
                     &           & Soft  &          &    1.35$\times 10^{-16}$  &          &                                            & 43.8        &           \\           
VLA J123713+621826   & \nodata   & Full  & 1.37     &    1.69$\times 10^{-15}$  &    2.4   & (1.1$\times 10^{22}$-1.1$\times 10^{23}$)  & (42.4-44.1) & 6.8-6.8   \\  
                     &           & Hard  &          &    1.19$\times 10^{-15}$  &          &                                            & (42.1-43.8) &           \\           
                     &           & Soft  &          &    4.71$\times 10^{-16}$  &          &                                            & (42.1-43.8) &           \\           
VLA J123721+621130   &    1.3    & Full  & 0.28     &    5.09$\times 10^{-16}$  &   10.8   &    1.9$\times 10^{23}$                     & 43.1        & 6.5       \\  
                     &           & Hard  &          &    5.07$\times 10^{-16}$  &          &                                            & 42.8        &           \\           
                     &           & Soft  &          &    4.45$\times 10^{-17}$  &          &                                            & 42.8        &           \\           
\enddata
\tablenotetext{a}{Alexander et al. assume a $\Gamma$ of 1.4 for
sources not detected in the hard band.  For those AGN with hard
band upper limits, however, we give the limit on $\Gamma$ as calculated
from the hard to soft band count rate ratio, $H/S$.}

\end{deluxetable}
\clearpage
\end{landscape}

\end{document}